\documentclass[journal=nalefd,manuscript=letter,english]{achemso}

\usepackage{graphicx}
\usepackage{color}
\usepackage[latin1]{inputenc}
\usepackage{amsbsy,amsmath,amssymb,amsthm,amsfonts}
\usepackage{times,mathptmx}
\usepackage{bm}
\usepackage{units}

\newcommand{\ik}{_{{\bf k}}}

\newcommand{\iq}{_{{\bf q}}}

\newcommand{\braket}[1]{\langle #1 \rangle}

\title{Microscopic view on the ultrafast photoluminescence from photo-excited graphene}

\author{Torben Winzer$^{1}$}
\author{Richard Ciesielski$^{2}$}
\author{Matthias Handloser$^{2}$}
\author{Alberto Comin$^{2}$}
\author{Achim Hartschuh$^{2}$}
\author{Ermin Malic$^{1}$}
\email{ermin.malic@tu-berlin.de}
\affiliation{$^{1}$Institut f\"ur Theoretische Physik, Technische Universit\"at Berlin,
  Hardenbergstr. 36, 10623 Berlin, Germany}
\affiliation{$^{2}$Department Chemie und CeNS, Ludwig Maximilians Universit\"at M\"unchen,  Butenandtstr. 5-13, 81377 Munich, Germany}

\begin{document}

\begin{abstract}
We present a joint theory-experiment study on ultrafast photoluminescence from photo-excited graphene. Based on a microscopic theory, we reveal two distinct mechanisms behind the occurring photoluminescence: Besides the well-known incoherent contribution driven by non-equilibrium carrier occupations, we found a coherent part that spectrally shifts with the excitation energy.
In our experiments, we demonstrate for the first time the predicted appearance and spectral shift of the coherent photoluminescence. 
\end{abstract}

The unique electronic band structure of graphene \cite{geim07} gives rise to distinct ultrafast phenomena, such as the appearance of a significant carrier multiplication \cite{winzer10, brida13, wendler14} and a transient optical gain \cite{li12, gierz13, winzer13}. While the ultrafast dynamics of optically excited carriers has been intensively investigated in literature 
\cite{dawlaty08, george08, sun08, plochocka09, wang10, breusing11, winnerl11, heinz13, gierz13, hofmann13, mit14, rana07-1, kim11, malic11-1}, there are only few experimental studies on the appearance of hot photoluminescence (PL) from graphene \cite{lui10, liu10, stoehr10, khanh13}.
A  broadband non-linear PL has been measured across the entire visible spectral range even exceeding the energy of the excitation laser. Its origin  has been traced back to  incoherent radiative recombination of thermalized carriers within simple theoretical models based on rate equations. 
A microscopic modeling of the PL including the competing processes of Coulomb-induced and phonon-assisted scattering as well as the impact of the coherence has not been performed, yet. 

\begin{figure}[t!]
  \begin{center}
\includegraphics[width=0.75\linewidth]{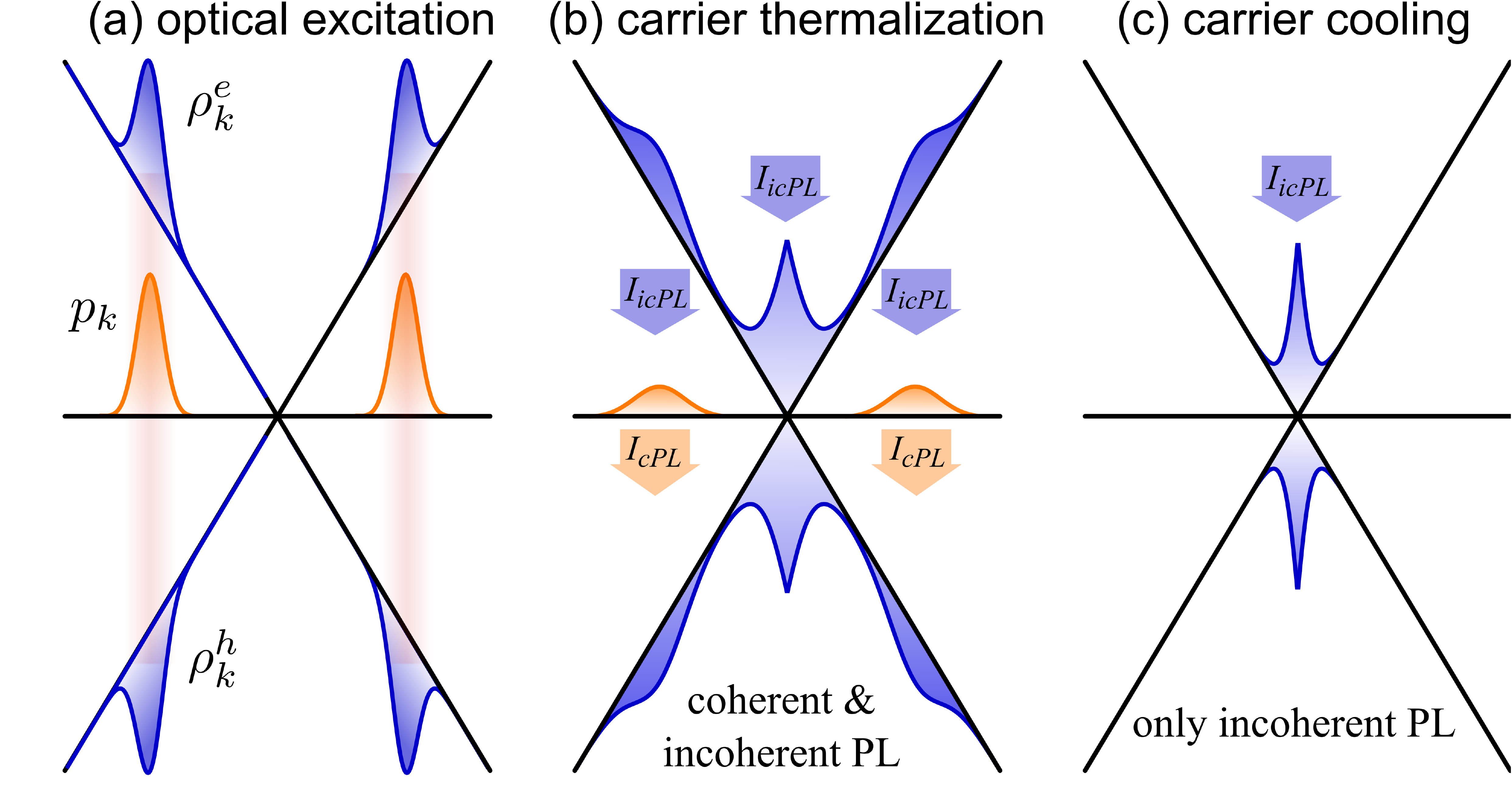}
  \end{center}
  \caption{Schematic illustration of the non-equilibrium dynamics in optically excited graphene. (a) Optical excitation (red shaded area) gives rise to a microscopic polarization $p_{\bf k}$ and a non-equilibrium carrier distribution $\rho_{\bf k}$. (b) Within the first tens of fs, the excited carriers thermalize into a spectrally broad Fermi distribution and the microscopic polarization is strongly damped via many-particle dephasing. (c) The carrier system cools down on a ps timescale resulting in a narrow distribution close to the Dirac point, while the polarization is already completely decayed. During the dynamics, radiative recombination of charge carriers occurs resulting in a hot photoluminescence (PL). Electron  and hole  occupations $\rho^e_{\bf k}$ and $\rho^h_{\bf k}$ drive the incoherent part  $I_{icPL}$, while the microscopic polarization $p_{\bf k}$ induces the coherent contribution $I_{cPL}$ that is always centered around the excitation energy. } \label{fig_scheme} 
\end{figure}

In this Letter, we present a  microscopic study offering access to spectrally and temporally resolved carrier and coherence dynamics as well as to the radiative recombination of excited carriers. 
We calculate the photoluminescence from photo-excited graphene and resolve two different microscopic mechanisms: the incoherent contribution driven by the non-equilibrium carrier and hole occupations and the coherent contribution induced by the microscopic polarization. The performed experiments demonstrate for the first time the appearance of the coherent PL by resolving its spectral shift with the excitation energy - in excellent agreement with our theoretical results. 

Optical excitation of graphene drives the microscopic polarization, which gives rise to non-equilibrium carrier distributions, cf. Fig. \ref{fig_scheme}(a). Coulomb-induced carrier-carrier scattering dominates the initial relaxation dynamics. The optically excited carriers scatter towards the Dirac point resulting in a hot thermalized distribution on a fs time scale. Depending on the excitation strength, this distribution can be spectrally very broad clearly exceeding the excitation energy, as illustrated in Fig. \ref{fig_scheme}(b). At the same time many-particle-induced dephasing damps and broadens the microscopic polarization. Carrier scattering with optical and acoustic phonons cools down the electronic system on a ps time scale accounting for a narrowing of the carrier distribution in the region close to the Dirac point (Fig. \ref{fig_scheme}(c)).  During the relaxation dynamics, the charge carriers can radiatively recombine resulting in a hot photoluminescence. Depending on the
  excitation regime, the PL can be also found well above the excitation energy, cf. Fig. \ref{fig_scheme}(b).
The intensity of the time-dependent photoluminescence is given by the photon flux \cite{kira99}
\begin{equation}
 \mathcal{I}({\bf q},t)\propto\frac{d}{dt}\sum\limits_{\mu}\langle c_{\bf q}^{\mu\dagger}c_{\bf q}^{\mu\phantom{\dagger}}\hspace{-3pt}\rangle(t),
\label{pl}
\end{equation}
where $c_{\bf q}^{\mu\dagger}$ and $c_{\bf q}^{\mu\phantom{\dagger}}$ create and annihilate a photon in the mode $\bf q$ with the polarization $\mu$, respectively. According to the experimental realization, we sum the PL over all polarizations. 
Exploiting the electron-photon Hamilton operator for interband transitions 
\begin{equation}
  H_{c-pt}=\sum\limits_{{\bf k},{\bf q},\mu}\left[\mathcal{M}_{{\bf k},{\bf q}}^{vc,\mu}\,a^{v\dagger}_{\bf k}a^{c\phantom{\dagger}}_{\bf k}c_{\bf q}^{\mu\dagger}+\mathcal{M}_{{\bf k},{\bf q}}^{cv,\mu}\,a^{c\dagger}_{\bf k}a^{v\phantom{\dagger}}_{\bf k}c_{\bf q}^{\mu\phantom{\dagger}}\right]
\end{equation}
we derive an equation of motion for the photon occupation $\langle c_{\bf q}^{\mu\dagger}c_{\bf q}^{\mu\phantom{\dagger}}\hspace{-3pt}\rangle$ yielding
\begin{equation}
\label{phass}
 i\hbar\frac{d}{dt}\braket{c_{{\bf q}}^{\mu\dagger}c_{{\bf q}}^{\mu\phantom{\dagger}}}=\sum\limits_{{\bf k}} \mathcal{M}_{{\bf k},{\bf q}}^{vc,\mu}\left[\braket{a^{v\dagger}_{{\bf k}}a^{c\phantom{\dagger}}_{{\bf k}}c_{{\bf q}}^{\mu\dagger}}-\braket{a^{c\dagger}_{{\bf k}}a^{v\phantom{\dagger}}_{{\bf k}}c_{{\bf q}}^{\mu\phantom{\dagger}}\hspace{-3pt}}\right]
\end{equation}
with the  creation and the annihilation operators for electrons $a_{\bf k}^{\dagger}$ and $a_{\bf k}^{\phantom{\dagger}}$,
the electron-photon matrix element $\mathcal{M}_{{\bf k},{\bf q}}^{\lambda\lambda',\mu}$, and the photon-assisted quantities $\braket{a^{\lambda\dagger}_{{\bf k}}a^{\lambda'\phantom{\dagger}}_{{\bf k}}c_{{\bf q}}^{\mu(\dagger)}}$. 
The matrix element determines the strength of the interaction and has the same symmetry as the semi-classical electron-field matrix element ${\bf{M}}_{{\bf k}}^{vc}$ \cite{malic13}, i.e. $\mathcal{M}_{{\bf k},{\bf q}}^{vc, \mu}\propto\hat{\bf e}_{\mu}\cdot{\bf M}_{{\bf k}}^{vc}$ with ${\bf{M}}_{{\bf k}}^{vc}=\langle v{\bf k} | \nabla| c{\bf k} \rangle$
and the unity vector $\hat{\bf e}_{\mu}$ denoting the polarization direction.
Next, we  derive a further equation of motion for the photon-assisted quantities appearing in Eq. (\ref{phass}). 
Neglecting the dynamics of coherent photons (i.e. $\braket{\dot{c}_{{\bf q}}^{\mu(\dagger)}}=0$)  and of stimulated emission that is expected to  marginally contribute to the photoluminescence at the considered excitation conditions, we obtain a closed system of equations
\begin{eqnarray}\label{photonassisted}
  i\hbar\frac{d}{dt}\braket{a^{v\dagger}_{{\bf k}}a^{c\phantom{\dagger}}_{{\bf k}}c_{{\bf q}}^{\mu\dagger}}(t)\hspace{-3pt}&=(\varepsilon\ik^c-\varepsilon\ik^v-\hbar\omega\iq)\braket{a^{v\dagger}_{{\bf k}}a^{c\phantom{\dagger}}_{{\bf k}}c_{{\bf q}}^{\mu\dagger}}(t)\\[5pt]\nonumber
&-\mathcal{M}_{{\bf k},{\bf q}}^{vc\mu}\Big[\rho\ik^e(t)\rho\ik^h(t)+p\ik(t) p^*\ik(t)\Big].
\end{eqnarray}
Here, $\rho\ik^c=\langle a_{\bf k}^{c\dagger}a_{\bf k}^{c\phantom{\dagger}}\hspace{-3pt}\rangle$  and $\rho\ik^h=1-\langle a_{\bf k}^{v\dagger}a_{\bf k}^{v\phantom{\dagger}}\hspace{-3pt}\rangle$ express the electron occupation in the conduction band and the hole occupation in the valence band, respectively.  Furthermore, the microscopic polarization $p\ik=\langle a_{\bf k}^{v\dagger}a_{\bf k}^{c\phantom{\dagger}}\hspace{-3pt}\rangle$ (often called coherence) expresses the probability for optically induced transitions between the valence and the conduction band. 
Within the Markov approximation \cite{kira99}, Eq. (\ref{photonassisted})   can be analytically solved allowing us to express  the experimentally accessible photoluminescence $I_{PL}({\bf q})$. It  corresponds to the photon flux from Eq. (\ref{pl}) integrated over the detection time $T$
\begin{eqnarray}\label{lum}
 I_{PL}({\bf q})\hspace{-3pt}&\hspace{-3pt}\propto \int_0^T 
\frac{d}{dt} \sum\limits_{\mu}\langle c_{\bf q}^{\mu\dagger}c_{\bf q}^{\mu\phantom{\dagger}}\hspace{-3pt}\rangle(t)
 \, dt=\frac{\pi}{\hbar}\int_0^T \sum\limits_{{\bf k}}|\mathcal{M}_{{\bf k},{\bf q}}^{vc}|^2 \hspace{-3pt}\times \\ \nonumber
& \Big[\underbrace{\rho\ik^e(t)\rho\ik^h(t)}_{\rightarrow \; I_{icPL}}+\underbrace{p\ik(t) p\ik^*(t)}_{\rightarrow \;I_{cPL}}\Big]\,
\delta(\varepsilon\ik^c-\varepsilon\ik^v-\hbar\omega\iq)\,dt\,.
\end{eqnarray}
\begin{figure}[t!]
\includegraphics[width=0.7\linewidth]{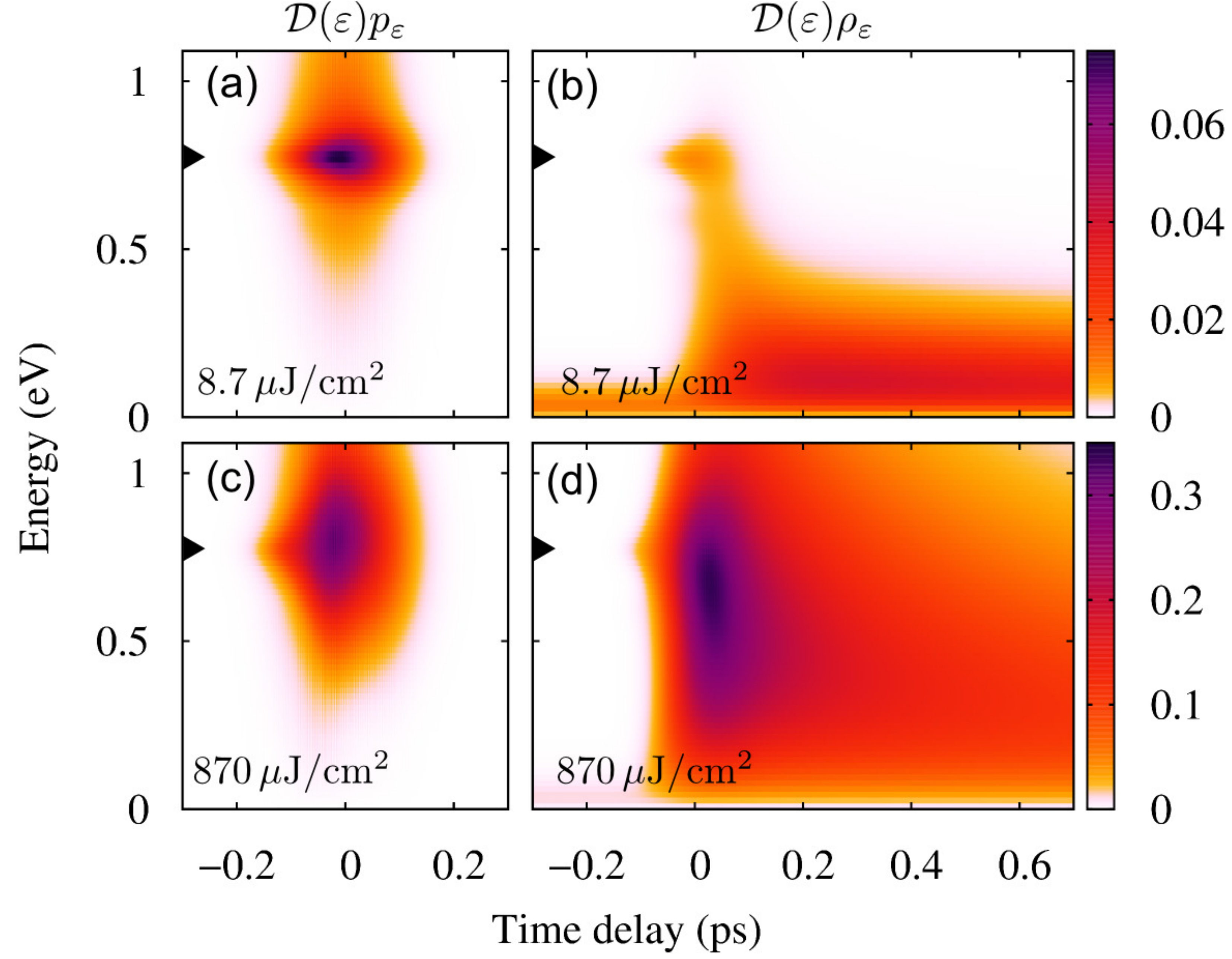}
   \caption{Surface plots illustrating the time- and energy-resolved dynamics of the microscopic polarization  $p_{\varepsilon}(t)$ [(a), (c)] and of the carrier occupation $\rho_{\varepsilon}(t)$ [(b), (d)] multiplied with the density of states $D(\varepsilon)$ at two distinct pump fluences representing a low- (upper figures) and a high-excitation regime (lower figures), respectively. While the polarization (coherence) is always centered at the excitation energy (black arrow), the carrier population can be spectrally very broad depending on the applied pump fluence.} \label{fig_rho_p} 
\end{figure}
We find that the photoluminescence has an incoherent term $I_{icPL}\propto \rho\ik^e (t) \rho\ik^h(t)$ driven by  the carrier and hole occupations and a coherent term $I_{cPL}\propto  p\ik(t) p\ik^*(t)$ driven by the square of the absolute value of the microscopic polarization.
The dynamics of these single-particle quantities is evaluated within the graphene Bloch equations \cite{malic13} -  a coupled set of differential equations for $p\ik(t)$, $\rho\ik^{\lambda}(t)$, and the phonon population $n\iq^j(t)$ for different optical and acoustic phonon modes $j$. Solving these equations within the second-order Born-Markov approximation \cite{malic13}, we obtain microscopic access to the ultrafast time- and energy-resolved Coulomb-induced and phonon-assisted carrier and coherence dynamics (Fig. \ref{fig_rho_p}) including the radiative recombination of excited carriers resulting in light emission from photo-excited graphene (Fig. \ref{fig_weak}).

Figure \ref{fig_rho_p} illustrates the time- and energy-resolved dynamics of the carrier occupation and the microscopic polarization for two different pump fluences representing a weak and a strong excitation regime, respectively. Our calculations reveal that independently of the applied pump fluence, the microscopic polarization is centered around the excitation energy and that it decays on a fs timescale, cf. Figs. \ref{fig_rho_p}(a) and (c). In contrast, the carrier occupation is strongly sensitive to the the pump fluence. The initial non-equilibrium distribution becomes thermalized due to carrier-carrier and carrier-phonon scattering on a fs time scale, where the pump fluence determines the spectral width of the carrier population. In particular, at high fluences electronic states far above the excitation energy are occupied, cf. Fig. \ref{fig_rho_p}(d). 

\begin{figure}[t!]
  \begin{center} 
\includegraphics[width=0.45\linewidth]{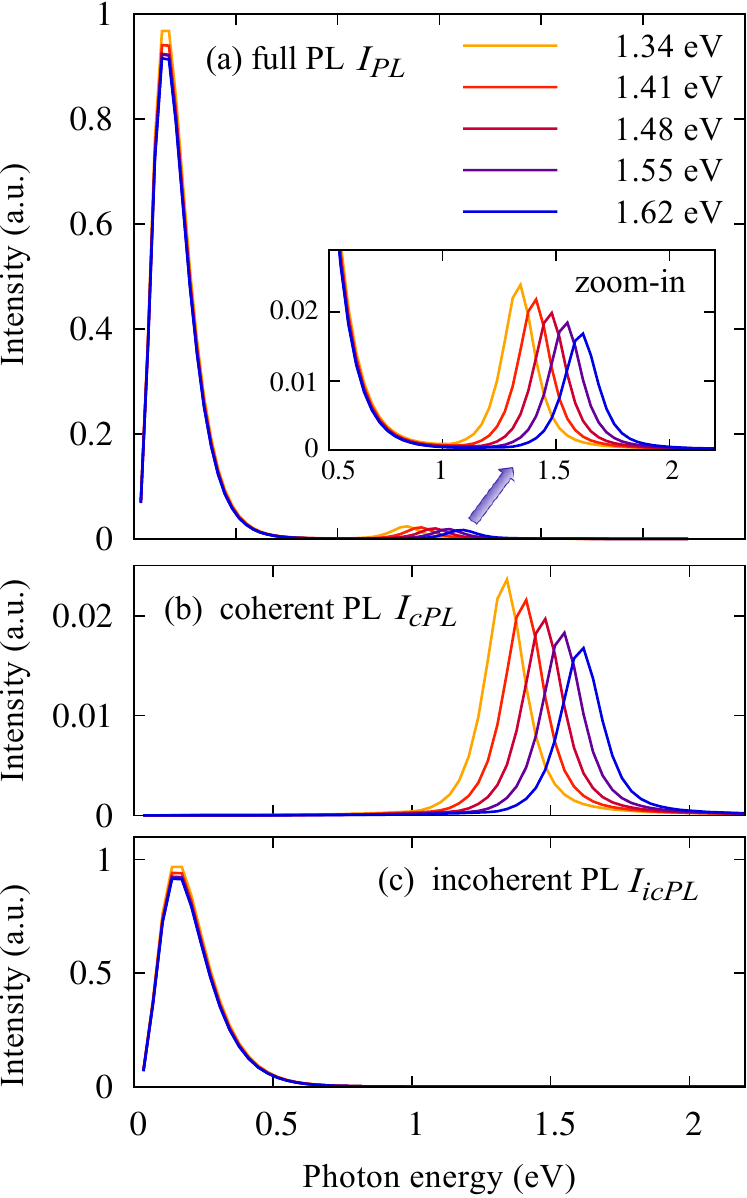}
  \end{center}
  \caption{(a) Photoluminescence emitted from optically excited graphene shown for varying excitation energies. The pump fluence is fixed to $\unit[8.7]{\mu J/cm^2}$ for all photon energies corresponding to a relatively weak excitation regime. The coherent ($I_{cPL}\propto p_{\bf k} p^*_{\bf k}$) and the incoherent ($I_{icPL}\propto\rho^c_{\bf k}\rho^h_{\bf k}$) contributions to the PL (cf. Eq. (\ref{lum})) are illustrated separately in (b) and (c). The first exhibits a clear shift with the excitation energy $\hbar \omega_L$, while the latter is centered at energies close to the Dirac point independently of $\hbar \omega_L$.} \label{fig_weak} 
\end{figure}
On the way to the equilibrium distribution, excited carriers can radiatively recombine giving rise to experimentally accessible photoluminescence.
Our calculations reveal that the photoluminescence depends on the excitation strength and the excitation energy.
We first analyze the impact of the latter at a constant relatively weak pump fluence, cf. Fig. \ref{fig_weak}(a). To obtain clear insights, we separately plot the coherent and the incoherent PL contributions (cf. Eq. (\ref{lum})) in Figs. \ref{fig_weak}(b) and (c), respectively. The coherent part $I_{cPL}$ is driven by the microscopic polarization $p_{\bf k}$ and it is centered around the excitation energy reflecting the behavior of $p_{\bf k}$, cf. Fig. \ref{fig_rho_p}(a). $I_{cPL}$ is characterized by a clear shift with the excitation energy. Since $p_{\bf k}$ decays on a fs timescale, the coherent PL stems from the initial dynamics during and immediately after the excitation pulse. In contrast, the incoherent part $I_{icPL}$ driven by non-equilibrium carrier occupations occurs during the entire carrier dynamics. The majority of thermalized carriers is energetically located below the applied excitation energy (Fig. \ref{fig_rho_p}(b)) and therefore, the maximum of $I_{icPL}$ is found close to the Dirac 
point, cf. Fig. \ref{fig_weak}(c). Remarkably, the incoherent spectrum is to a large extent independent of the excitation energy, since the thermalized carrier distribution is only determined by the applied pump fluence.

\begin{figure}[t!]
\includegraphics[width=0.5\linewidth]{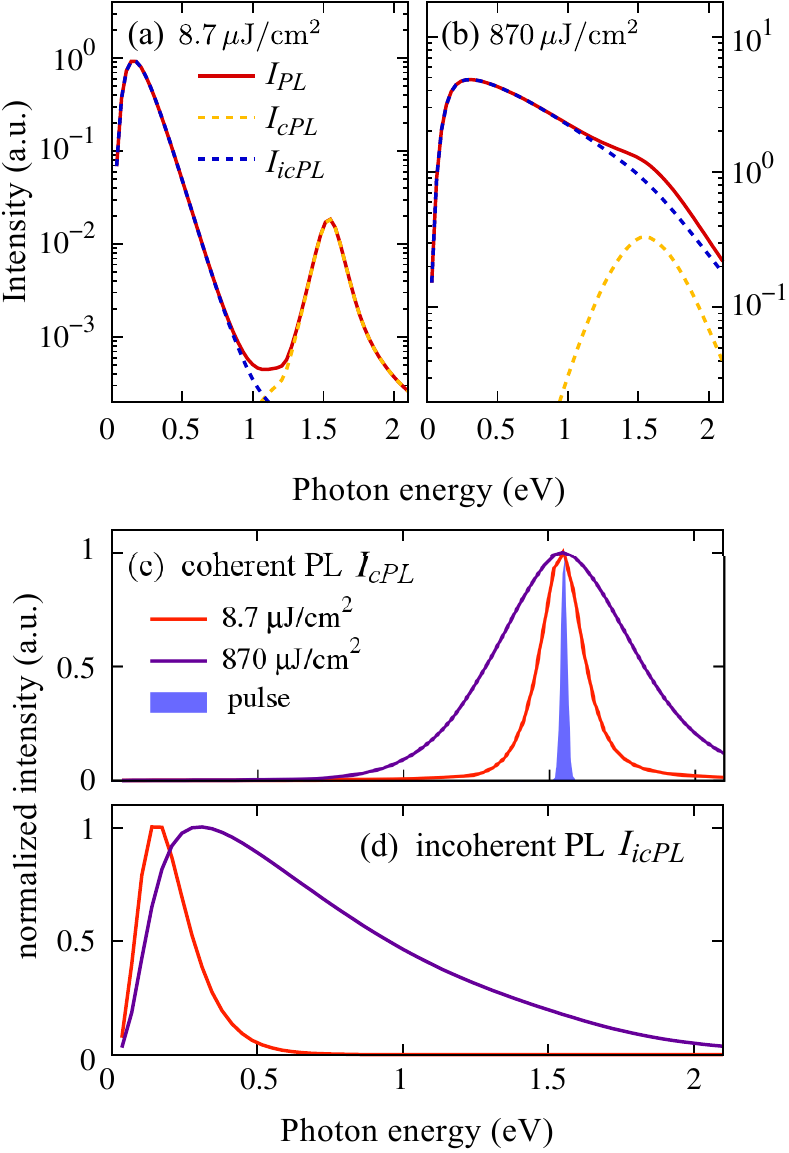}
  \caption{Illustration of the incoherent $I_{icPL}$ and the coherent photoluminescence $I_{cPL}$ in the (a) weak  and  (b) strong excitation regime represented by pump fluences of $\unit[8.7]{\mu J/cm^2}$ and $\unit[870]{\mu J/cm^2}$, respectively. The excitation energy is set to \unit[1.55]{eV}. For the weak excitation, both contributions are spectrally well separated, while in the strong excitation case the incoherent part predominates the entire spectral range. The fluence-dependent spectral width of the two normalized PL contributions is shown separately in (c) and (d). For better comparison, the intensities are normalized. The blue shaded area in (c) indicates the optical excitation pulse.} \label{fig_sep} 
\end{figure}

Since the pump fluence has a crucial impact on the photoluminescence, we display a direct comparison between the PL in the low- and in the high-excitation regime, cf.  Fig. \ref{fig_sep}.
For weak excitations, the coherent and the incoherent contributions  $I_{cPL}$ and $I_{icPL}$ give rise to two well separated peaks in the logarithmic plot of the PL, cf. Fig. \ref{fig_sep}(a). In contrast, for strong excitations, the thermalized carrier occupation corresponds to a spectrally broad Fermi distribution 
with a significant number of carriers also in the range of the excitation energy. These carriers decay much slower compared to the dephasing of the spectrally narrow microscopic polarization (Fig. \ref{fig_rho_p}). As a result, the incoherent contribution $I_{icPL}$ dominates the radiative recombination in the strong excitation regime leading to pronounced PL in a broad spectral range including energies well above the excitation. To further illustrate the impact of the pump fluence on the spectral width of the PL, the two PL contributions are shown separately in Figs. \ref{fig_sep} (c) and (d). Compared to the optical excitation pulse (blue shaded area), the coherent PL exhibits a significantly larger spectral broadening due to the many-particle-induced dephasing (Fig. \ref{fig_sep}(c)). The spectral range of the incoherent PL is even more sensitive to the pump fluence. In the lower excitation regime, the incoherent contribution remains centered below $\unit[0.5]{eV}$, whereas for stronger excitations a 
significant PL is observed also above the excitation energy of $\unit[1.55]{eV}$. Furthermore, the maximum peak shows a blue-shift with the increasing pump fluence reflecting the spectral shift of the maximum of the thermalized carrier density (Fig. \ref{fig_rho_p}). 

So far, only the spectrally broad incoherent part of the photoluminescence has
been reported in the literature~\cite{lui10,liu10,stoehr10}.  
The coherent part is difficult to be measured, since it is centered around the
excitation energy, where optical filters are needed to protect the
detector. Here, we perform experiments measuring the PL in the region close to
the excitation energy allowing us to identify the coherent contribution to the
PL. Emission spectra were recorded for few layer graphene using
a scanning confocal microscope with NA = 1.3. The spectral
and temporal characteristics of the broadband laser excitation
pulse provided by a Ti:S oscillator were controlled using a 4f
pulse shaper based on a dual liquid crystal line array (cf. Figs.~S1
and Figs.~S2 of the supplementary material). 
The emission spectra were carefully corrected for background signal 
contributions from residual laser scattered light, the sample substrate, and
the microscope objective recorded directly next to
the graphene flake (cf. Fig.~S3 of the supplementary material). The spectra
were also normalized by the sensitivity and transmission of the detection
setup, measured separately using a calibrated black body radiation source.

\begin{figure}[t!]
\includegraphics[width=0.5\linewidth]{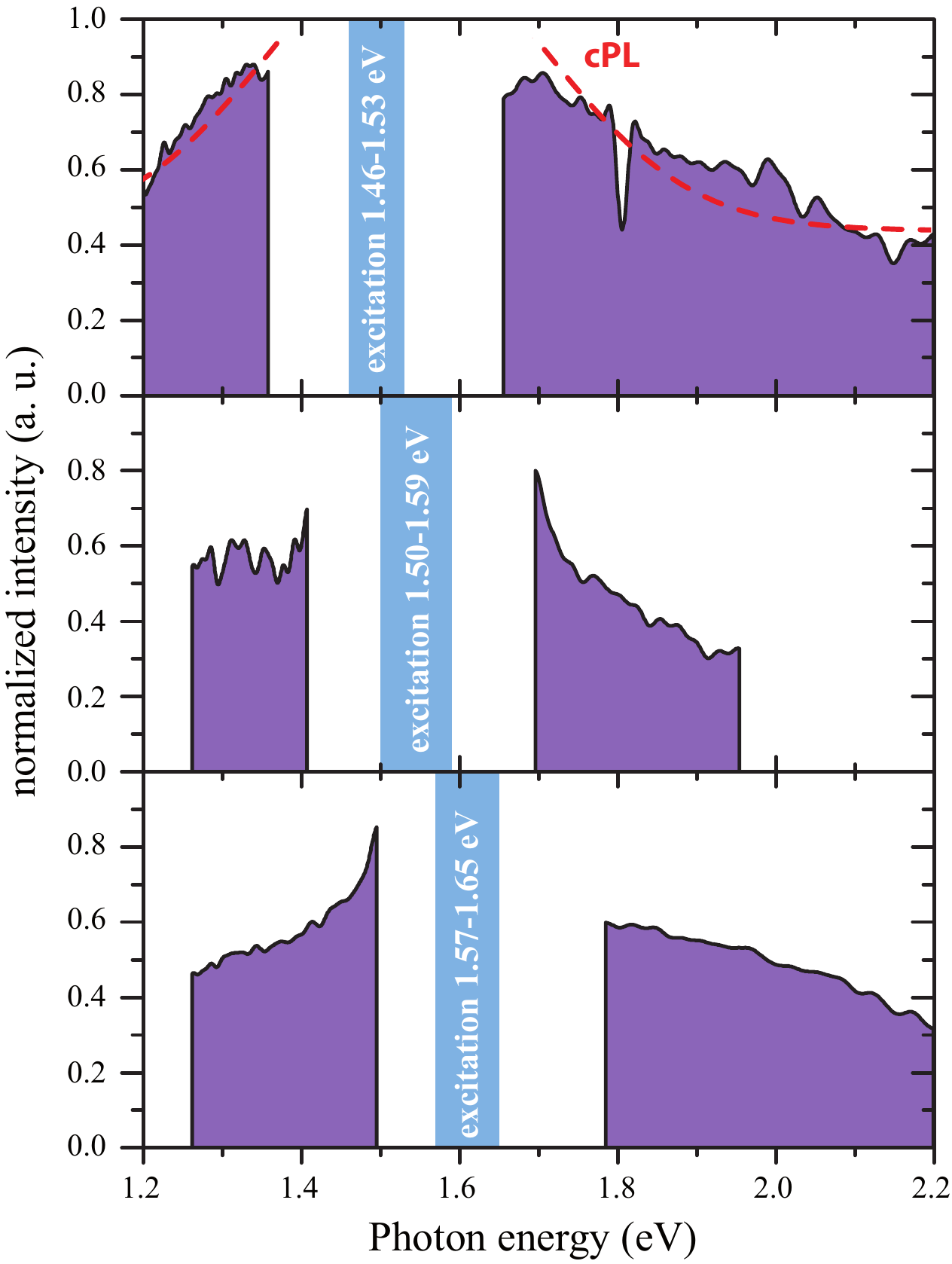}
  \caption{
  Experimental photoluminescence spectra of few layer
graphene demonstrating the coherent part of the PL. The experiments
are recorded on the blue and on the red side of the excitation pulse
for three different excitation energies as indicated. Laser scattered
light was blocked using appropriate spectral filters. In addition to the
broad incoherent contribution that is predicted to continuously decay
towards higher photon energies the PL intensity is seen to increase towards
the excitation pulse energy from both sides, the hallmark of the 
coherent polarization contribution (pump fluence \unit[320]{$\mu$ J/cm$^2$}). Furthermore, in the upper figure, we included an extrapolated Voigt profile assuming a spectrally flat incoherent PL contribution in the detected range (as guide to the eye). The full width at half maximum of the peak function is about 0.4 eV that is comparable to the theoretical values.
  } \label{fig_exp} 
\end{figure}

Figure 5 shows the emission spectra detected on the blue and the red
  side of the excitation pulse for three different excitation energies.
For the experiment, the laser excitation energies and the detection ranges
were selected to exclude signal contributions from four-wave
mixing~\cite{mikhailov10}.
This near-degenerate signal increases rapidly towards
  the excitation energy and could thus be mistaken for the coherent
  polarization signal. On the other hand, four-wave mixing can be excluded by
  limiting the detection ranges such that $\omega_{detected} \neq \omega_3 = 2 \omega_1 -\omega_2$ for all  $\omega_{1,2}$ within the excitation window, cf. Fig.~S2 of the supplementary material.
While this allows us to observe the wings of the coherent polarization
contribution only, the detected signal clearly exhibits the features predicted
by our microscopic theory:
The PL intensity increases towards the excitation pulse energy from the red
and the blue side and its maximum shifts with the excitation energy.
The incoherent photoluminescence that is predicted to continuously decay
towards higher photon energies is seen as an additional
broad signal contribution that does not show major differences in all three spectra. In the shown energy range, the incoherent PL does
not reach zero even far away from the excitation window. 

We have also measured the fluence dependence of the emission spectrum, cf. Fig. S4 of the supplementary material.
In the strong excitation regime, we find a pronounced low-energy contribution
that is constant over the measurable spectral range. 
In contrast, in the weak excitation regime, the emission has the opposite
behavior and increases towards the excitation energy. These observations
corroborate the theory results presented in Figs.~\ref{fig_sep}(a)-(b).

In summary, we have presented a joint theory-experiment study on the ultrafast
photoluminescence in photo-excited graphene. For the first time, we have
identified a coherent contribution to the photoluminescence that is
characterized by a clear shift with the excitation energy. Our study gives
microscopic insights into the elementary processes behind the carrier and
photon dynamics in graphene. \\

We acknowledge financial support from the Deutsche Forschungsgemeinschaft
through SPP 1459 and 1391 and the ERC through the starting Grant NEWNANOSPEC. Sample support was provided by Antonio Lombardo and Andrea Ferrari. Furthermore, we thank A. Knorr for inspiring discussions.


\providecommand*\mcitethebibliography{\thebibliography}
\csname @ifundefined\endcsname{endmcitethebibliography}
  {\let\endmcitethebibliography\endthebibliography}{}

\end{document}